\documentclass[11 pt]{article}
\usepackage[mathscr]{eucal}
\usepackage{amssymb,latexsym}
\usepackage{verbatim}
\usepackage{amsmath}
\usepackage{amsthm}
\usepackage{enumerate}
\usepackage{authblk}
\usepackage{color}
\usepackage{multicol}
\usepackage{tikz}
\usepackage{xypic}
\usepackage{caption}
\usepackage{cite}
\usepackage{hyperref}
\usetikzlibrary{snakes,decorations.pathmorphing}
\usepackage{subcaption}

\usepackage[nottoc,notlot,notlof]{tocbibind}

\definecolor{bello}{RGB}{82,74,208}

\newtheorem{thm}{Theorem}[section]
\newtheorem{Def}[thm]{Definition}

\hypersetup{
    colorlinks=true,
    linkcolor=blue,
    filecolor=magenta,      
    urlcolor=cyan,
    citecolor=black,
}

\newcommand\wt{\widetilde}

\renewcommand\d{\partial}
\newcommand\e{\varepsilon}
\renewcommand\b{\beta}

\renewcommand\div{{\rm div}}

\newcommand\g{\gamma}

\renewcommand\a{\alpha}

\renewcommand\t{\tau}

\newcommand\beq{\begin{equation}}
\newcommand\eeq{\end{equation}}
\newcommand\ben{\begin{enumerate}}
\newcommand\een{\end{enumerate}}
\newcommand\bit{\begin{itemize}}
\newcommand\eit{\end{itemize}}

\DeclareMathOperator{\Ric}{Ric}

\DeclareMathOperator{\diver}{div}

\DeclareMathOperator{\Hess}{Hess}

\renewcommand{\div}{\diver}

\newcommand{\R}{\mathbb R}

\newcommand{\ov}{\overline}

\newcommand{\ext}{\text{{\rm ext}}}

\newcommand{\pd}{\partial}

\newcommand{\mc}{\mathcal}
\newcommand{\mf}{\mathfrak}

\def\undertilde#1{\mathord{\vtop{\ialign{##\crcr
   $\hfil\displaystyle{#1}\hfil$\crcr\noalign{\kern1.5pt\nointerlineskip}
   $\hfil\tilde{}\hfil$\crcr\noalign{\kern1.5pt}}}}}

\newcounter{mnotecount}

\title{Anisotropic examples of inflation-generating initial conditions for the big bang}

\author[1]{Eric Ling\footnote{el@math.ku.dk. 
}}
\author[2]{Annachiara Piubello \footnote{annachiara.piubello@uni-potsdam.de. Research supported by the DFG Project ME 3816/3-1, part of the SPP2026.}}

\affil[1]{Copenhagen Centre for Geometry and Topology (GeoTop),
\linebreak
Department of Mathematical Sciences, University of Copenhagen, Denmark}

\affil[2]{Institute of Mathematics, University of Potsdam, Germany. }

\begin{document}
\date{}
\maketitle
\vspace{.0in}

\begin{abstract}
The inflationary scenario, which states that the early universe underwent a brief but dramatic period of accelerated spatial expansion, has become the current paradigm of early universe cosmology. Although inflationary cosmology has its many successes, it does not (as of yet) have the status of an established physical theory. In this paper, we provide mathematical support for the inflationary scenario in a class of anisotropic spacetimes by generalizing the work in \cite{Ling_remarks_cosmo}. These anisotropic spacetimes satisfy certain initial conditions so that they are perfectly isotropic at the big bang but become less isotropic as time progresses. The resulting inflationary eras are a consequence of the initial conditions which force the energy-momentum tensor to be dominated by a cosmological constant at the big bang.
 
\end{abstract}



\tableofcontents

\medskip
\medskip

\newpage

\section{Introduction}

\subsection{Cosmic inflation}

The inflationary scenario has become the current paradigm of early universe cosmology. Roughly, it states the following.

\medskip

\noindent{\bf The inflationary scenario:} In the early universe, before the radiation-dominated era, there was a brief but dramatic period of accelerated spatial expansion.

\medskip

The inflationary scenario  was proposed in the late 1970s and early 80s \cite{Guth:1980zm,Starobinsky:1980te,Linde:1981mu} as a solution to some problems in the standard big bang model, e.g., the flatness and horizon problems. It was soon realized that inflation can provide a framework for generating the seeds of the large-scale structures in our universe \cite{Mukhanov:1981xt}. Observations of the anisotropies in the CMB radiation performed by COBE, WMAP, and most recently by Planck\cite{Planck:2018jri} support these claims.

Given the many successes of the inflationary scenario, it is perhaps not too surprising that most papers on early universe cosmology give the impression that inflation has been firmly established and observationally proven. However there are many inflationary models that can be in agreement with observation \cite{Martin:2013tda}. In fact, any theory which predicts an almost flat universe with a nearly scale-invariant curvature power spectrum, small tensor-to-scalar ratio, and small Gaussian fluctuations would be in agreement with current data,  e.g., \cite{Brandenberger_bouncing}.

Moreover, although phenomenologically successful, current realizations of inflationary models suffer from conceptual problems, perhaps none more so than the problem of initial conditions \cite{Brandenberger:2016uzh, Linde_2018}. In fact there are conflicting opinions on the naturalness of initial conditions for inflation \cite{Ijjas:2013vea, Guth_2014}.

Most papers on initial conditions for inflation begin in an inhomogeneous universe with an energy-momentum tensor dominated by an inflaton scalar field in a slow-roll potential and see if the resulting dynamics can produce an inflationary era followed by a homogeneous universe.  This is not our approach. Our approach is purely geometrical. Quantities of interest are described solely in terms of a special unit timelike vector field $u$ (whose integral curves represent the comoving observers in the universe) and the spacetime metric $g$. 

In this paper we provide mathematical support for the inflationary scenario. In section~\ref{sec-ex}, we show that inflation arises for a class of anisotropic spacetimes from special geometrical initial conditions. Our initial conditions are stated informally in section \ref{sec: geom initial} and formally in section \ref{sec: main thm}. These anisotropic spacetimes  are examples of our main result, Theorem \ref{thm-main}, which concludes from the special initial conditions that the Ricci tensor (and hence also the energy-momentum tensor) is dominated by a cosmological constant at the big bang.
 Theorem \ref{thm-main} is a generalization of the main result in \cite{Ling_remarks_cosmo}. In fact, a major inspiration for this paper was to find anisotropic examples of the main result in \cite{Ling_remarks_cosmo}. 

The benefit of our geometrical approach is its conceptual clarity: we will describe precisely which comoving observers experience inflation and how fast they are accelerating solely in terms of the unit timelike vector field $u$ and spacetime metric $g$.

Our geometrical initial conditions can be thought of as a certain type of fine-tuning condition for the big bang. As briefly reviewed in the next section, a Boltzmannian viewpoint on the arrow of time suggests that some type of fine-tuning initial condition for the big bang should exist.

\subsection{Inflation and the arrow of time}\label{sec: inflation and the arrow of time}

An obvious feature of our universe is the existence of an arrow of time. We observe certain processes in our everyday experience, but we hardly ever observe those same processes time-reversed. A vase shatters into a multitude of pieces, but we never observe these pieces spontaneously arranging themselves perfectly together into a vase. The second law of thermodynamics is postulated to explain the arrow of time, and a modern Boltzmannian mindset of the second law leads to the conclusion that the universe began with special, non-generic, fine-tuned initial conditions.

It was Penrose who originally argued \cite{Penrose_time} that the overall arrow of time we observe is linked to special initial conditions for the universe that are drastically far away from the dynamical trend towards gravitational collapse. He calculated that the entropy of the radiation-dominated early universe is around 30 orders of magnitude smaller than the Beckenstein-Hawking entropy of its corresponding black hole state. See also \cite{albrecht2003cosmic, carroll2004spontaneous}. 

With this understanding, the homogeneous and isotropic assumptions of the standard FLRW models of cosmology are a reasonable choice of initial conditions as they match exceedingly well with current observations. However, some inflationary cosmologists instinctively take a different perspective. They seek an explanation for the large-scale isotropy of the universe from dynamical processes during inflation. But special initial conditions  -- by their nature -- go against dynamical trends. That is, the creation of special initial conditions from dynamics beginning with generic initial conditions seems contradictory. Summarizing, some inflationary cosmologists seek generic initial conditions for the universe, but those who adopt a Boltzmannian point of view of the second law of thermodynamics (as we do) argue that initial conditions should instead be special in order to explain the arrow of time. 

 While inflationary theory alone may not suffice to explain the large-scale isotropy of the universe, it still has many successes and remains the prevailing paradigm in early universe cosmology. The simplest way to generate inflation is to introduce an inflaton scalar field in a slow-roll potential -- a methodology that is somewhat adhoc since it simply postulates the existence of a scalar field for which we have no direct evidence for. So a natural inquiry is to ask if there is other evidence to support the inflationary scenario. A primary motivation of this paper is to demonstrate that there is mathematical evidence in support of the inflationary scenario. We will see that inflation is inevitable provided certain  geometrical initial conditions are assumed at the big bang, and, as discussed in this section, some degree of fine-tuning in the initial conditions is anticipated.

\subsection{Geometrical initial conditions for the big bang}\label{sec: geom initial}

In this section we describe, informally, the primary geometrical initial conditions we will be considering in our main result, Theorem \ref{thm-main}. These initial conditions are supposed to mimic -- without assuming isotropy -- the geometrical properties at the big bang. 

Let's first clarify what we mean by the ``big bang" as there are conflicting view points in the literature. For us the big bang refers to a time when the scale factor limits to zero. For example, if the scale factor is $a(\tau) = \tau$ (as in the Milne model), then the big bang corresponds to $\tau = 0$. If the scale factor is $a(\tau) = e^\tau$ (as in the flat de Sitter model), then the big bang corresponds to $\tau = -\infty$.

To motivate the type of geometrical initial conditions we will be considering, we focus on scale factor perturbations of the Milne model, which have been dubbed ``Milne-like spacetimes" in \cite{Galloway_Ling_inextend}. These models were extensively studied in \cite{Ling_coord_singu}, detailing possible applications to fundamental problems in cosmology. See also \cite{Ling_Piubello, Nomura:2021lzz,PhysRevD.21.3305}. They are $k = - 1$ FLRW spacetimes whose scale factor satisfies $a(\tau) \approx \tau$ for $\tau$ near $\tau = 0$. (An inflating example would be $a(\tau) = \sinh(\tau)$.) Interestingly, for Milne-like spacetimes, the big bang appears as a coordinate singularity, and so they extend into a larger spacetime. 

\begin{figure}[h]
\[
 \begin{tikzpicture}[scale = 0.65]

\shadedraw [color = white, bottom color=bello!40!white, top color = white] (-8.5,2) -- (-8.5,-2.12) -- (0.5,-2.12) -- (0.5,2);


\draw (-4.35,2.5) node [scale = .75] {$\tau$};


	\draw (1.5, -2) node [scale = .75] {$\tau = 0$};
	\draw [thick, red] (-8.5,-1) -- (0.5,-1);
	\draw (2.2,-1) node [scale = .75] {$\tau = \text{constant}$  };


	\draw [thick, black] (-4 + 1.125,-1.95) -- (-4 + 1.125,2);
	\draw [thick, black] (-4 + 2*1.125,-2) -- (-4 + 2*1.125,2);
	\draw [thick, black] (-4 + 3*1.125,-2.01) -- (-4 + 3*1.125,2);
	\draw [thick, black] (-4 - 1.125,-1.905) -- (-4 - 1.125,2);
	\draw [thick, black] (-4 - 2*1.125,-1.93) -- (-4 - 2*1.125,2);
	\draw [thick, black] (-4 - 3*1.125,-1.95) -- (-4 - 3*1.125,2);
	\draw [thick, black] (-4,-2) -- (-4 ,2);
	
	
	\draw[snake=zigzag, red, thick] (-8.46,-2) -- (0.5,-2);
\draw[snake=zigzag, white, line width=1.5mm] (-8.46,-2.2) -- (0.5,-2.2);  

\draw (-4,-4.5) node [scale = .75] {$g \,=\, -d\tau^2 + a^2(\tau)h$ };
\draw [<->,thick] (-4,-3.5) -- (-4,2.25);



\shadedraw [color = white, bottom color=bello!40!white, top color = white]  (6,2) -- (10,-2) -- (14,2);
\draw [dashed, thick, blue] (10,-2) -- (14,2);
\draw [dashed, thick, blue] (10,-2) -- (6,2);

\draw [<->,thick] (10,-3.5) -- (10,2.25);
\draw [<->,thick] (5.5,-2) -- (14.5,-2);

\draw (9.65,2.5) node [scale = .75] {$t$};
\draw (14.75, -2.25) node [scale = .75] {$x^i$};

\draw (9.75,-2.25) node [scale = .75] {$\mc{O}$};


	\draw [thick, red] (6.5,2) .. controls (10, -1.3).. (13.5,2);
	\draw [->] [thick]  (13.9,0.2) arc [start angle=-90, end angle=-155, radius=40pt];
	\draw (15.4,.2) node [scale = .75] {$\tau = \text{constant}$ };


	\draw [thick, black] (10,-2) -- (12,2);
	\draw [thick, black] (10,-2) -- (13,2);
	\draw [thick, black] (10,-2) -- (11,2);
	\draw [thick, black] (10,-2) -- (9,2);
	\draw [thick, black] (10,-2) -- (8,2);
	\draw [thick, black] (10,-2) -- (7,2);
	\draw [thick, black] (10,-2) -- (10,2);

\draw (10,-4.5) node [scale = .75] {$g \,=\, \Omega^2\big(\tau)[-dt^2 + dx^2 + dy^2 + dz^2]$ };

\end{tikzpicture}
\]
\captionsetup{format=hang}
\caption{\small{A Milne-like spacetime represented in two different coordinate systems. On the left, standard comoving coordinates are used; the metric is degenerate at $\tau = 0$. On the right, conformal Minkowskian coordinates are used; the metric is nondegenerate at $\tau = 0$ which corresponds to the lightcone at $\mc{O}$. The black lines depict the comoving observers.}}
\label{fig: comoving observers}
\end{figure}
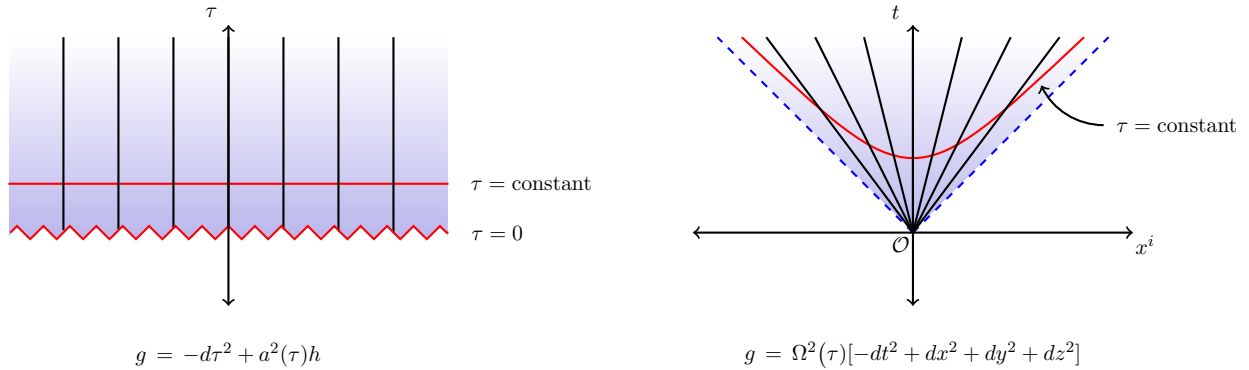

Recall that the comoving observers are the integral curves of the vector field $u$ given by $u = \pd_\tau$ in comoving coordinates. As illustrated in Figure \ref{fig: comoving observers}, the comoving observers for a Milne-like spacetime all emanate from a single point $\mc{O}$ in the extended spacetime, which is just the origin $(0,0,0,0)$ in the conformal Minkowskian coordinates $(t,x,y,z)$. We refer to this property as ``$\mc{O}$ being an origin point for $u$," see Definition \ref{def: Origin point}. The existence of an origin point $\mc{O}$ for $u$ is a highly fine-tuned and non-generic assumption. Recall that some fine-tuning is to be expected from the discussion on the arrow of time in section  \ref{sec: inflation and the arrow of time}.

An origin point $\mc{O}$ for $u$ is the first main assumption in Theorem \ref{thm-main}. The other main assumption is that the energy-momentum tensor $T$ approaches that of a perfect fluid at $\mc{O}$. See Definition \ref{def: perfect fluid limit}.  This assumption is more physically convincing than assuming that $T$ is exactly a perfect fluid (as in the FLRW models) since we expect small deviations from perfect isotropy in our universe.
 Therefore the perspective taken here is that the universe began in a state of perfect isotropy at the big bang. This is the crux of Definition \ref{def: perfect fluid limit}. Moreover, this perspective is reinforced in our examples since the shear vanishes towards the big bang, see eq. \eqref{eq- shear to 0}.

An ``origin point $\mc{O}$ for $u$" and ``$T$ approaching a perfect fluid at $\mc{O}$" are the two primary assumptions in Theorem \ref{thm-main}. There are three other assumptions that are purely technical. The conclusion of Theorem \ref{thm-main} is that the energy-momentum tensor is precisely given by a cosmological constant at $\mc{O}$. This fact will be used in section \ref{sec: inflation scenario} to prove the existence of inflationary eras in our anisotropic examples.

\newpage

\newpage

\section{The main theorem}\label{sec: main thm}

The initial conditions stated informally in section \ref{sec: geom initial} will be stated formally in this section. Our main result, Theorem \ref{thm-main}, is a generalization of the main result (Theorem 2.2) in \cite{Ling_remarks_cosmo}. Anisotropic examples of our main theorem are provided in section \ref{sec: the examples}, and we prove the existence of inflationary eras for these examples in section \ref{sec: inflation scenario}.

We set our conventions. Our definition of a spacetime $(M,g)$ will follow \cite{Ling_causal_theory}. (Except that, for simplicity, we will assume that all spacetimes are four-dimensional.) The manifold $M$ is always assumed to be smooth. A $C^k$ spacetime is one where the metric $g$ is $C^k$, that is, its components $g_{\mu\nu} = g(\pd_\mu, \pd_\nu)$ are $C^k$ functions with respect to any coordinates $(x^0, \dotsc, x^3)$.  A \emph{continuous} spacetime is one where the metric is continuous, that is, its components are continuous functions with respect to any coordinates. Our definitions of timelike curves and the timelike future $I^+$ will also follow \cite{Ling_causal_theory}.

Let $(M,g)$ be a $C^k$ spacetime. A $C^0$ spacetime $(M_\ext, g_\ext)$ is said to be a \emph{continuous spacetime extension} of $(M,g)$ provided there is an isometric embedding 
\[
(M,g) \:\: \hookrightarrow \:\: (M_\ext, g_\ext)
\]
preserving time orientations such that $M \subset M_\ext$ is a proper subset. ($M$ is in fact an open submanifold of $M_\ext$ since they are both four-dimensional.) Note that we are identifying $M$ with its image under the embedding. We remark that $g_\ext$ is $C^2$ in the examples constructed in the next section.

\medskip
\begin{Def}[Origin point]\label{def: Origin point}
\emph{
Let $(M_\ext, g_\ext)$ be a continuous spacetime extension of a $C^k$ spacetime $(M,g)$. Let $u$ be a unit future directed timelike vector field on $M$. We say that a point {\bf $\boldsymbol{\mc{O}}$ \emph{is an  origin point for}} $\boldsymbol{u}$ if $\mc{O} \in M_\ext \setminus M$ and $\mc{O}$ is a past endpoint for each integral curve of $u$, and each extended integral curve is $C^1$ at $\mc{O}$. (Clearly this implies $\mc{O}$ lies in the closure $\ov{M}$ within $M_\ext$.) In other words, $\mc{O}$ is an origin point for $u$ if each integral curve of $u$, parameterized as $\g \colon (0,b) \to M$, satisfies 
\begin{itemize}
\item[(i)]$\displaystyle\lim_{\t \to 0}\g(\t) = \mc{O}$,
\item[(ii)] $\wt{\g}'(0)$ exists and $\wt{\g}'(0) = \displaystyle\lim_{\t \to 0} \g'(\t)$, 
\end{itemize}
where $\wt{\g} \colon [0,b) \to M_\ext$ is the extended curve defined by $\wt{\g}(0) = \mc{O}$ and $\wt{\g}(\t) = \g(\t)$ for $\t > 0$. Continuity of the metric implies $\wt{\g}'(0)$ is a unit future directed timelike vector.
}
\end{Def}

\smallskip

\noindent\emph{Remarks.} Definition \ref{def: Origin point} is supposed to model the behavior of the comoving observers in Figure \ref{fig: comoving observers} (right). It is essentially the same as assumption (b) in \cite[Thm. 2.2]{Ling_remarks_cosmo}. Actually, Definition \ref{def: Origin point} is slightly stronger; we assume this stronger assumption since it's easier to state and all the examples in section \ref{sec-ex} will satisfy it.

\medskip

We recall some terminology from section 2 of \cite{Ling_remarks_cosmo}. Let $\mc{O} \in M_\ext \setminus M$ be an origin point for $u$. A $C^k$ function $f \colon M \to \R$ \emph{extends continuously} to $M \cup \{\mc{O}\}$ if there is a continuous function $\wt{f} \colon M \cup \{\mc{O}\} \to \R$ such that $\wt{f}|_M = f$. In this case, we call $\wt{f}$ the \emph{continuous extension} of $f$. 
A $C^k$ tensor $T$ defined on $M$ \emph{extends continuously} to $M \cup \{\mc{O}\}$ if there is a coordinate neighborhood $U$ of $\mc{O}$ with coordinates $(x^0, \dotsc, x^3)$ such that each of the components of $T$ extends continuously to $(U \cap M) \cup \{\mc{O}\}$. (This definition does not depend on the choice of coordinate system by the usual transformation law for tensor components.) This defines a continuous tensor $\wt{T}$ on $M \cup \{\mc{O}\}$, called the \emph{continuous extension} of $T$, which satisfies $\wt{T}|_M = T$. For example, the metric tensor $g$ extends continuously to $M \cup \{\mc{O}\}$ (by definition of a continuous extension). Trivially, if $T$ is a smooth tensor defined on all of $M_\ext$, then clearly $T|_M$ extends continuously to $M \cup \{\mc{O}\}$.

\begin{Def}[Limiting to a perfect fluid near $\mc{O}$]\label{def: perfect fluid limit}
\emph{
Let $(M,g)$ be a $C^2$ spacetime, and let $\mc{O} \in M_\ext \setminus M$ be an origin point for $u$. Let $T$ be the energy-momentum tensor on $M$ (i.e., $T = \frac{1}{8\pi}G$ in suitable units where $G = \text{Ric} - \frac{1}{2}Rg$ is the Einstein tensor). Let $\rho_0, p_0 \in \R$. We say that $\boldsymbol{T}$ {\bf\emph{limits to a perfect fluid $\boldsymbol{(u, \rho_0, p_0)}$ at} $\boldsymbol{\mc{O}}$} if 
\begin{itemize}
\item[(i)] $\rho := T(u,u)$ extends continuously to $M \cup \{\mc{O}\}$ and $\wt{\rho}(\mc{O}) = \rho_0$,
\item[(ii)] for any unit spacelike vector field $e$ on $M$, which is orthogonal to $u$, the function $p_e := T(e,e)$ extends continuously to $M \cup \{\mc{O}\}$ and $\wt{p}_e(\mc{O}) = p_0$,
\item[(iii)] $T - T_{\rm perfect}$ extends continuously to $M \cup \{\mc{O}\}$ and its continuous extension is zero at $\mc{O}$, where $T_{\rm perfect}$ is the tensor on $M$ given by
\[
T_{\rm perfect} \,=\, (\rho_0 + p_0)u_* \otimes u_* + p_0g,
\] 
where $u_* = g(u,\cdot)$ is the one-form metrically equivalent to $u$.
\end{itemize}
}
\end{Def}

\medskip

\noindent\emph{Remark.} Definition \ref{def: perfect fluid limit} relaxes the requirement that $T$ is identically a perfect fluid in assumption (a) of \cite[Thm. 2.2]{Ling_remarks_cosmo}. Moreover, it's  more physically convincing:
FLRW models have perfect fluid energy-momentum tensors, and we expect that an FLRW model approximates our universe better as we go back in time towards the big bang.

\medskip

Lastly, we require a mild, technical timelike convexity assumption:

\begin{Def}[Locally timelike convex near $\mc{O}$]\label{def: locally timelike convex}
\emph{
Let $\mc{O} \in M_\ext \setminus M$ be an origin point for $u$. Let $\g \colon (0,b) \to M$ be an integral curve of $u$. We say {\bf $\boldsymbol{M}$ \emph{is locally timelike convex about} $\boldsymbol{\g}$ \emph{near} $\boldsymbol{\mc{O}}$} if there is an $\e > 0$ and a coordinate neighborhood $U\subset M_\ext$ centered at $\mc{O}$ with coordinates $(x^0, \dotsc, x^3)$ satisfying 
\begin{itemize}
\item[(i)] $g_{\mu\nu}(\mc{O}) = \eta_{\mu\nu}$ and $|g_{\mu\nu}(p) - \eta_{\mu\nu}| < \e$ for all $p \in U$,
\item[(ii)] $\pd_0|_\mc{O} = \wt{\g}'(0)$,
\item[(iii)] $I^+_{\eta_{\e}}(\mc{O}, U) \subset M$,
\end{itemize}
where $\eta_{\e}$ is the narrow Minkowskian metric given by $\eta_{\e} = -\frac{\e}{2-\e}(dx^0)^2 + \delta_{ij}dx^idx^j$.
  } 
\end{Def}

\smallskip

\noindent\emph{Remarks.} In \cite[Thm. 2.2]{Ling_remarks_cosmo}, it was assumed that the manifold $M$ satisfies $M = I^+(\mc{O}, M_\ext)$. Definition \ref{def: locally timelike convex} relaxes this requirement and is a much weaker assumption. It will hold for the examples constructed in section \ref{sec-ex}. 
Also, conditions (i) and (ii) in Definition \ref{def: locally timelike convex} will always be satisfied by continuity of the metric and applying the Gram-Schmidt orthogonalization process appropriately. The heart of Definition \ref{def: locally timelike convex} is condition (iii)  and is the motivation for the terminology ``timelike convex near $\mc{O}$." 

\medskip

We are now ready to state our main theorem which generalizes \cite[Thm. 2.2]{Ling_remarks_cosmo}.

\smallskip

\begin{thm}\label{thm-main}
Let $(M_\ext, g_\ext)$ be a continuous spacetime extension of a $C^2$ spacetime $(M,g)$. Let $u$ be a unit future directed timelike vector field on $M$. Assume the following.
\begin{itemize}
\item[\emph{(a)}] $\mc{O} \in M_\ext \setminus M$ is an origin point for $u$.
\item[\emph{(b)}] The energy-momentum tensor $T$ on $M$ limits to a perfect fluid $(u, \rho_0, p_0)$ at $\mc{O}$.
\item[\emph{(c)}] $M$ is locally timelike convex about $\g$ near $\mc{O}$ for some integral curve $\g$ of $u$. 
\item[\emph{(d)}] The Ricci tensor $\emph{Ric}$ on $M$ extends continuously to $M \cup \{\mc{O}\}$.
\item[\emph{(e)}] $(M_\ext, g_\ext)$ is strongly causal at $\mc{O}$.
\end{itemize}
Then
\[
\rho_0 \,=\, -p_0.
\]
Moreover, the continuous extension of \emph{Ric} at $\mc{O}$ is given by
\[
\wt{\Ric}|_{\mc{O}} \,=\, 8\pi \rho_0\,g_\ext|_{\mc{O}}.
\]
\end{thm}

\medskip
\noindent\emph{Remark.} Assumptions (a), (b), and (c) are Definitions \ref{def: Origin point},  \ref{def: perfect fluid limit}, and \ref{def: locally timelike convex}, respectively. Assumption (d) will be satisfied whenever $(M_\ext, g_\ext)$ is a $C^2$ extension of $(M,g)$, which is the case for the examples constructed in the next section. Assumption (e) is a technical assumption needed for the proof; it's satisfied, for example, whenever $M_\ext$ is a subset of a globally hyperbolic spacetime.

\medskip

\proof
Seeking a contradiction, assume $\rho_0 \neq -p_0$. Then 
\begin{align*}
u_* \otimes u_* \,&=\, \frac{1}{\rho_0 + p_0}(T_{\rm perfect} - p_0g)
\\
&=\, \frac{1}{\rho_0 + p_0}\big((T_{\rm perfect} - T) + T - p_0g\big).
\end{align*}
By assumption (b),  $T_{\rm perfect} - T$ extends continuously to $M \cup \{\mc{O}\}$, and its continuous \linebreak extension is zero at $\mc{O}$. Also $T$ extends continuously to $M \cup \{\mc{O}\}$ by assumption (d). Therefore $u_* \otimes u_*$ extends continuously to $M \cup \{\mc{O}\}$. As in the proof of \cite[Thm. 2.2]{Ling_remarks_cosmo}, this implies that the vector field $u$ extends continuously to $M \cup \{\mc{O}\}$. However, assumptions (c) and (e) prove that $u$ does \emph{not} extend continuously. Heuristically, this can be seen in Figure~\ref{fig: comoving observers} (right). Rigorously, this follows from an analogous contradiction argument used in the proof of \cite[Thm. 2.2]{Ling_remarks_cosmo}. 
Thus we have  $\rho_0 = -p_0$.

Next we prove that $\wt{\text{Ric}}|_{\mc{O}} = 8\pi \rho_0 \,g_\ext|_{\mc{O}}$. The Einstein equations imply
\begin{align*}
\text{Ric} \,&=\, 8\pi T + \frac{1}{2}R g
\\
&=\, 8\pi (T - T_{\rm perfect}) + 8\pi T_{\rm perfect} + \frac{1}{2}Rg
\\
&=\, 8\pi (T - T_{\rm perfect}) + 8\pi T_{\rm perfect} - 4\pi(\text{tr}\,T)g.
\end{align*}
Since $\rho_0 = -p_0$, we have $T_{\rm perfect} = -\rho_0 g$, and so $T_{\rm perfect}$ extends continuously to $M \cup \{\mc{O}\}$. Also $\text{tr}\,T$ extends continuously to $M \cup \{\mc{O}\}$, and its continuous extension is $-\rho_0 + 3p_0 = - 4\rho_0$ at $\mc{O}$. Therefore evaluating the above expression at $\mc{O}$ gives
\[
\pushQED{\qed}
\wt{\text{Ric}}|_{\mc{O}} \,=\, 0 - 8\pi\rho_0\, g_\ext|_{\mc{O}} + 16\pi \rho_0\, g_\ext|_{\mc{O}} \,=\, 8\pi \rho_0 \,g_\ext|_{\mc{O}}. \qedhere
\popQED
\]

\section{Anisotropic examples of the main theorem}\label{sec-ex}

In section \ref{sec: the examples} we construct explicit examples of spacetimes satisfying the hypotheses of Theorem \ref{thm-main}. Clearly any Milne-like spacetime with a $C^2$ spacetime extension will satisfy the hypotheses of the theorem. But the goal of this section is to construct anisotropic examples as well, i.e., examples that are not FLRW spacetimes. (Recall Milne-like spacetimes are $k = -1$ FLRW spacetimes and hence are isotropic.) Briefly, to achieve this, we generalize Milne-like spacetimes in the following way: In spherical coordinates $(t,r,\theta,\varphi)$, the comoving observers in a Milne-like spacetime are parameterized by the curves $t = \mu r$ for $1 < \mu \leq \infty$, see Figure \ref{fig: comoving observers}. ($\mu = \infty$ corresponds to the comoving observer traveling along $r = 0$.) In our anisotropic examples, we stipulate that the comoving observers follow the trajectories $t = \mu f(r)$, where $f(r) \approx r$ for $r$ small. Like Milne-like spacetimes, the metric is still conformally flat and the conformal factor is a function of the foliation of the spacelike hypersurfaces orthogonal to the comoving observers, i.e., the conformal factor is a function of the rest spaces of $u$.

In section \ref{sec: inflation scenario}, we use the conclusion of Theorem \ref{thm-main} (that the Ricci tensor, and hence also the energy-momentum tensor, is dominated by a cosmological constant) to show that those comoving observers with $\mu$-value greater than some critical number $\mu_{\rm crit}$ will experience inflationary eras, lending support to the inflationary scenario. Our analysis depends on investigating the terms in the Raychaudhuri equation as they approach the origin point $\mc{O}$.

\subsection{The examples}\label{sec: the examples}

In this section we construct explicit examples of spacetimes satisfying the hypotheses of Theorem \ref{thm-main}. Our examples will depend on only two functions $f(r)$ and $\Phi(\zeta)$.

Let $f(r)$ be a smooth positive function on $[0,\infty)$ satisfying $f(r) = r + O(r^3)$ as $r \to 0$ and $f'(r)\geq 1$ for all $r\geq 0$.\footnote{We have $f(0) = 0$, $f'(0) = 1$, and $f''(0) = 0$. In fact  $f(r) = r + O_2(r^3)$.} A simple example of such a function is $f(r) = \sinh(r)$. Our manifold of interest is
\begin{equation}\label{eq: M def 1}
M \,:=\, \lbrace (t,x,y,z) \mid  t >  f(r),  \text{ where } r = \sqrt{x^2 + y^2 + z^2}\rbrace,
\end{equation}
equipped with the metric
\begin{equation}\label{eq: M def 2}
g \,=\, e^{2\Phi(\zeta)} [-dt^2+dx^2+dy^2+dz^2]
\end{equation}
for some arbitrary smooth\footnote{More generally $\Phi(\zeta)$ only needs to be a $C^2$ function on a neighborhood of $\zeta = 0$, but, for simplicity, we assume $\Phi(\zeta)$ is smooth on all of $\R$. } function $\Phi(\zeta)$ on $\R$. Here $\zeta = \zeta(t,r)$ is given by
\begin{equation}\label{eq: M def 3}
\zeta(t,r) = \frac{t^2}{2} - \int_0^r \frac{f(s)}{f'(s)}\,ds. 
\end{equation}

The spacetime extension $(M_\ext, g_\ext)$ of $(M,g)$ is simply defined by extending $g$ to all $\R^4 \approx M_\ext$. In fact the metric is $C^2$ on $M_\ext$, which follows from the assumptions on $f(r)$.

\medskip
\medskip
\noindent\emph{Remark.} The simple case $f(r) = r$ corresponds to (a subclass of) Milne-like spacetimes  \cite{Ling_coord_singu}. This follows since the conformal factor is a function of $t^2 - r^2$.
\medskip

\begin{figure}
  \centering
  \begin{subfigure}[b]{0.49\textwidth}
    \[
 \begin{tikzpicture}[scale = .7]
\shadedraw [color = white, bottom color=bello!40!white, top color = white] (5.7,2.3) -- (10,-2) -- (14.3,2.3);
\draw [dashed, thick, blue] (10,-2) -- (14,2);
\draw [dashed, thick, blue] (10,-2) -- (6,2);
 \draw(9.75,-2.3) node [scale = .75] {$\mc{O}$};

\draw [<->,thick] (10,-3.5) -- (10,2.25);
\draw [<->,thick] (5.5,-2) -- (14.5,-2);

\draw (9.65,2.5) node [scale = .75] {$t$};
\draw (14.75, -2.25) node [scale = .75] {$x^i$};


	\draw [thick,red] (6.5,2) .. controls (10, -1.3).. (13.5,2);
	\draw [->] [thick]  (13.9,0.2) arc [start angle=-90, end angle=-155, radius=40pt];
	\draw (15.4,.2) node [scale = .75] {$\tau = \text{constant}$ };

        \draw [color = black](11.5,2.3) node [scale = .75] {$t = \mu r $};
        \draw [thick, color = black] (10,-2) -- (13,2);
	\draw [thick, color = black] (10,-2) -- (11.5,2);
	\draw [thick, color = black] (10,-2) -- (8.5,2);
	\draw [thick, color = black] (10,-2) -- (7,2);
	\draw [thick, color = black] (10,-2) -- (10,2);
\draw (10,-4.5) node [scale = .85] {$M := \lbrace (t,x,y,z) \mid  t >  r\rbrace, \quad g = e^{2\Phi(\tau(t,r))}\eta$};
\draw (10,-5.5) node [scale = .85] {with $\tau(t,r) = \frac{t^2}{2} - \frac{r^2}{2}. $};

\end{tikzpicture}
\]
  \end{subfigure}
  \hfill
  \begin{subfigure}[b]{0.49\textwidth}
    \[
 \begin{tikzpicture}[scale = .7]
\shadedraw [color = white, bottom color=bello!40!white, top color = white] (6.7,2.2).. controls (7.4,1) .. (10,-2) .. controls (12.6,1) .. (13.3,2.2);
\draw [dashed, thick, blue] (10,-2) .. controls (12.6,1) .. (13.3,2.2);
\draw [dashed, thick, blue] (10,-2)  .. controls (7.4,1) .. (6.7,2.2);
 \draw(9.75,-2.3) node [scale = .75] {$\mc{O}$};

\draw [<->,thick] (10,-3.5) -- (10,2.25);
\draw [<->,thick] (5.5,-2) -- (14.5,-2);

\draw (9.65,2.5) node [scale = .75] {$t$};
\draw (14.75, -2.25) node [scale = .75] {$x^i$};


	\draw [thick, red] (7,2) .. controls (10, -1.5).. (13,2);
	\draw [->] [thick]  (13.8,0.2) arc [start angle=-90, end angle=-155, radius=46pt];
	\draw (15.4,.2) node [scale = .75] {$u$-orthogonal };

    \draw [color = black](11.5,2.3) node [scale = .75] {$t = \mu f(r) $};
	\draw [thick, color = black] plot [smooth] coordinates {(10,-2) (10.9, -.75) (11.65, .75) (12,2)};
        \draw [thick, color = black] plot [smooth] coordinates {(10,-2) (10.43, -.73) (11,2)};
        \draw [thick, color = black] (10,-2) -- (10,2);	
        \draw [thick, color = black] plot [smooth] coordinates {(10,-2) (9.57, -.73) (9,2)};
        \draw [thick, color = black] plot [smooth] coordinates {(10,-2) (9.1, -.75) (8.35, .75) (8,2)};

\draw (10,-4.5) node [scale = .85] {$M := \lbrace (t,x,y,z) \mid  t >  f(r)\rbrace, \quad g = e^{2\Phi(\zeta(t,r))} \eta$};
\draw (10,-5.5) node [scale = .85] {with $\zeta(t,r) = \frac{t^2}{2} - \int_0^r \frac{f(s)}{f'(s)}\,ds. $};

\end{tikzpicture}
\]
  \end{subfigure}
  \caption{On the left is a Milne-like spacetime represented in conformal Minkowskian coordinates. On the right, the anisotropic examples constructed in this section. They are constructed to look like a Milne-like spacetime around the origin $\mc{O}$. The comoving observers (i.e., the integral curves of $u$) still emanate from the origin, and the manifold is still foliated by slices orthogonal to the comoving observers.}
  \label{fig:both}
\end{figure}
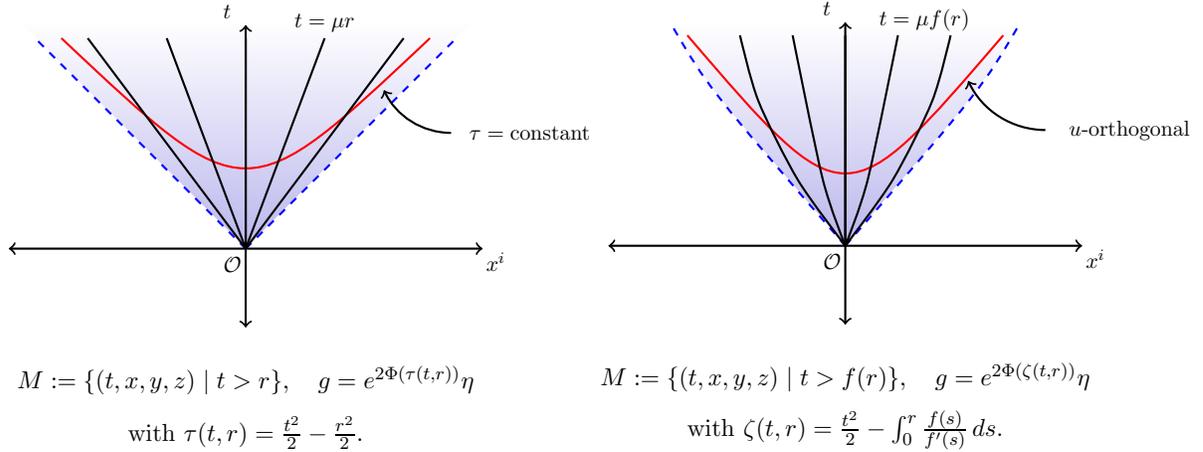

The unit future directed timelike vector field $u$ (whose integral curves are the comoving observers) will be given by normalized gradient of $\zeta$:
\begin{equation}\label{eq: u}
u \,:=\,-\frac{\nabla \zeta}{|\nabla \zeta|_g} \,=\, \frac{e^{-\Phi(\zeta)}}{\sqrt{ t^2 f'(r)^2 -f(r)^2}}  \Big(t \, f'(r) \, \d_t +  f(r) \d_r\Big).
\end{equation}

By construction the integral curves of $u$ emanate from the origin $\mc{O} = (0, 0,0,0)$ in $(t,x,y,z)$-coordinates. Each integral curve of $u$ follows the trajectory of the curve $t = \mu f(r)$ for $1 < \mu \leq \infty$ (with $\mu = \infty$ corresponding to $r = 0$). To see this, recognize that the curve
\begin{equation}\label{eq-integral curve}
r \,\mapsto\, \big(\mu f(r), r, \theta_0 ,\varphi_0\big)
\end{equation}
  in $(t,r,\theta, \varphi)$-coordinates has tangent vector parallel to $u$. By rewriting these curves in $(t,x,y,z)$-coordinates, it's clear that they extend as $C^1$ curves through the origin $\mc{O}$. Thus $\mc{O}$ is an origin point for $u$. Hence part (a) of Theorem \ref{thm-main} is verified.


Now we verify properties (b) through (e) of Theorem \ref{thm-main}. Property (c) is evidently satisfied; simply consider the integral curve along the $t$-axis given by $r = 0$. Property (e)  holds since $M_\ext$ is conformal to Minkowski spacetime. Property (d) holds since the metric is $C^2$ on all of $M_\ext$.

 The remainder of this section will be dedicated to proving property (b), namely, that the energy-momentum tensor converges to that of a perfect fluid. However, to gain control over the terms appearing in the energy-momentum tensor, we found it easier to work with the following subset of our original manifold:
\begin{equation}\label{eq: M epsilon def}
M_\e \,:=\, \lbrace (t,x,y,z) \mid  t > (1 + \e) f(r),  \text{ where } r = \sqrt{x^2 + y^2 + z^2}\rbrace,
\end{equation}
where $\e > 0$ is arbitrary. Note that $M_\e$ approaches $M$ as $\e \to 0$. Moreover, for any $\e > 0$, we see that $M_\e$ also satisfies properties (a) and (c) - (e) of Theorem \ref{thm-main}.

Now we prove property (b) of Theorem \ref{thm-main} with $M_\e$ playing the role of $M$ in statement of Theorem \ref{thm-main}. And this will hold for any $\e > 0$. The following fact will be used.
 
 \medskip
 \medskip
 
\noindent\underline{Fact:} We have the following bound on $M_\e$:

\begin{equation}\label{eq-limit}
    \left|\frac{t f(r)  }{t^2 f'(r)^2 - f(r)^2 }\right| \,<\, \frac{1+\e}{2\e + \e^2}\:.
\end{equation}


\medskip

\noindent\emph{Proof of fact.}
Since $M_\e$ is only defined for $t>(1+\e)f(r)$, we have

\begin{align*}
    \left|\frac{t f(r)  }{t^2 f'(r)^2 - f(r)^2 }\right| \,&=\, \frac{f(r)  }{t f'(r)^2} \left(1 + \frac{f(r)^2}{t^2 f'(r)^2 - f(r)^2 } \right) \\
    & <\, \frac{1  }{(1+\e)f'(r)^2} \left(1 + \frac{1}{(1+\e)^2 f'(r)^2 -1} \right)
    \\
    &\leq \, \frac{1}{1+\e}\left(1 + \frac{1}{(1+\e)^2  -1} \right),
\end{align*}
where we used the positivity of $f(r)$ and the fact that $f'(r)\geq 1$. \qed
\\

  We start by showing property (i) in Definition \ref{def: perfect fluid limit}. From conformal geometry, the Ricci tensor is given by
\begin{equation}\label{eq-conformal}
    R_{\a\b} = - 2(\Hess \Phi)_{\a\b} + 2\nabla_\a\Phi\nabla_\b\Phi-\big(\square\Phi+2|d\Phi|_\eta ^2\big)\eta_{\a\b}.
\end{equation}
Here, all operators on the right-hand side are taken with respect to the Minkowski metric $\eta$. Using \eqref{eq-conformal}, we have
\begin{align*}
    8\pi T_{\a\b} &= R_{\a\b} - \frac12 Rg_{\a\b}\\
    &= -2 (\Hess \Phi)_{\a\b} + 2\nabla_\a \Phi \nabla_\b \Phi + \square \Phi\, \eta_{\a\b} + 2|d\Phi|^2_\eta\, \eta_{\a\b},
\end{align*}
Straightforward computation shows 
\begin{align}
\rho &= T(u,u) \notag \\
    &\label{eq-bad limit}  = \frac{e^{-2 \Phi(\zeta)}}{8\pi}  \left[ \left(2 +\frac{4f(r)}{r f'(r)}
  - \frac{2  t^2 f(r)  f''(r)}{ t^2 f'(r)^2-f(r)^2 }\right)\Phi'(\zeta) +\frac{3(t^2 f'(r)^2 -f(r)^2)}{f'(r)^2 }  \Phi'(\zeta)^2\right].
\end{align}

We are interested in showing that $\rho$ extends continuously to $M_\e\cup \lbrace\mc{O}\rbrace$ and finding its limit at the origin $\mc{O}$. Both the first and second terms will contribute to $\wt{\rho}(\mc{O})$ since, for small $r$, we have:
$$2+\frac{4f(r)}{r f'(r)} = 6 + O(r^2).$$
Additionally, by utilizing \eqref{eq-limit}, we see that the third term vanishes at the origin $\mc{O}$. Finally, the fourth term in \eqref{eq-bad limit} also vanishes since $f'(r)=1+O(r^2)$.

Hence, $\rho$ extends continously to the origin and
\begin{align}\label{eq-rho origin}
    \wt{\rho}(\mc{O}) \,=\, \frac{3}{4\pi} e^{-2 \Phi(0)}  \Phi'(0).
\end{align}
This shows property (i) in Definition \ref{def: perfect fluid limit}. To show (ii), consider the vector field
$$v \,=\, \frac{e^{-\Phi(\zeta)}}{\sqrt{t^2 f'(r)^2 -f(r)^2 }}  \Big(f(r) \, \d_t +t \, f'(r) \d_r\Big).$$
 By construction $v$ is unit spacelike and orthogonal to $u$. Let $e_\theta,e_\varphi$ be the standard orthonormal vectors on the sphere so that $\lbrace u,v,e_\theta,e_\varphi\rbrace$ forms an orthonormal basis on $M_\e$ (modulo some spherical coordinate singularities). Straightforward computations show
\begin{equation*}
    \begin{split}
    p_v \,&=\, T(v,v) \\
    & = \frac{e^{-2 \Phi(\zeta)}}{8\pi}  \left[ \left(-2 -\frac{4f(r)}{r f'(r)}
  - \frac{2   f(r)^3  f''(r)}{f'(r)^2 (t^2 f'(r)^2-f(r)^2 )}\right)\Phi'(\zeta)\right.\\ &\,\left.\hspace{2.1cm} -\frac{t^2 f'(r)^2 -f(r)^2}{f'(r)^2 }  (2\Phi''(\zeta)+
    \Phi'(\zeta)^2)\right],\\
    p_{e_\theta} & =\, T(e_\theta,e_\theta)\\
    & =\,  \frac{e^{-2 \Phi(\zeta)}}{8\pi }  \left[ \left(-4 -\frac{2f(r)}{r f'(r)}
    + \frac{2 f(r)  f''(r)}{f'(r)^2}\right)\Phi'(\zeta) -\frac{t^2 f'(r)^2 -f(r)^2}{f'(r)^2 }  (2\Phi''(\zeta)+  \Phi'(\zeta)^2)\right],\\
    p_{e_\varphi} & = \, p_{e_\theta} .
\end{split}
\end{equation*}
Using \eqref{eq-limit} again, we see that these functions extend continuously to $M_\e \cup \{\mc{O}\}$ and 
\begin{align*}
    \wt{p}_v(\mc{O}) \,=\, \wt{p}_{e_\theta}(\mc{O}) \,=\, \wt{p}_{e_\varphi}(\mc{O}) \,=\, -\frac{3}{4\pi} e^{-2 \Phi(0)}  \Phi'(0).
\end{align*}
To finish our analysis, we need to compute all the cross terms of $T$. These cross terms are 
\begin{equation*}
\begin{split}
    T(u,v) &= -\frac{2 e^{-2 \Phi (\zeta)}}{8\pi}\frac{ t f(r)^2  f''(r)}{f'(r)\big(t^2f'(r)^2 - f(r)^2\big)}\Phi'(\zeta), \\
    T(u,e_\theta) &= T(u,e_\varphi) = T(v,e_\theta) = T(v,e_\varphi) = T(e_\theta,e_\varphi)=0.
\end{split}
\end{equation*}
Moreover, using \eqref{eq-limit}, we have
\begin{equation}\label{eq-limit Tuv}
    T(u,v) \to 0.
\end{equation}
as we approach $\mc{O}$ within $M_\e$. 

Let $e_0,e_1,e_2,e_3$ denote $u,v,e_\theta,e_\varphi$ respectively. If $e$ is any unit spacelike vector field orthogonal to $u$, then it can be written as $e=\sum_{i=1}^3 a_i e_i$ with $\sum_{i=1}^3 a_i^2 = 1$. Then vanishing of the cross terms implies
$$p_e  = T(e,e) = \sum_{i=1}^3 a_i^2 \,T(e_i,e_i), $$
and so in the limit
\begin{equation}\label{eq-p origin}
    \wt{p}_e(\mc{O}) = -\frac{3}{4\pi} e^{-2 \Phi(0)}  \Phi'(0).
\end{equation}
It is only left to show property (iii) in Definition \ref{def: perfect fluid limit}, i.e., that $T- T_{\rm perfect}$ extends continuously to $M_\e \cup \{\mc{O}\}$ and is zero at $\mc{O}$. Recall that 
\[
T_{\rm perfect} \,=\, (\rho_0 + p_0)u_* \otimes u_* + p_0g,
\]
where $\rho_0$ and $p_0$ are given by \eqref{eq-rho origin} and \eqref{eq-p origin}. 

We work in $(t,x,y,z)$-coordinates as they clearly cover the origin $\mc{O}$. 
We have
\begin{equation*}
\begin{split}
    \d_t &= \frac{e^{\Phi(\zeta)}}{\sqrt{t^2 f'(r)^2-f(r)^2}} \big(tf'(r) u +f(r)v\big) \\
    \d_x &= e^{\Phi(\zeta)}\left[\frac{\sin(\theta)\cos(\varphi)}{\sqrt{t^2 f'(r)^2-f(r)^2}} \big(tf'(r) u +f(r)v\big) +\cos(\theta)\cos(\varphi) e_\theta -\sin(\varphi)e_\varphi \right]\\
    \d_y &= e^{\Phi(\zeta)}\left[\frac{\sin(\theta)\sin(\varphi)}{\sqrt{t^2 f'(r)^2-f(r)^2}} \big(tf'(r) u +f(r)v\big) +\cos(\theta)\sin(\varphi) e_\theta +\cos(\varphi)e_\varphi \right]\\
    \d_z &= e^{\Phi(\zeta)}\left[\frac{\cos(\theta)}{\sqrt{t^2 f'(r)^2-f(r)^2}} \big(tf'(r) u +f(r)v\big) -\sin(\theta) e_\theta \right].\\
\end{split} 
\end{equation*}
Then 
\begin{equation*}
\begin{split}
    T(\d_t,\d_t) &= \frac{e^{2\Phi(\zeta)}}{t^2 f'(r)^2-f(r)^2} \big[ t^2f'(r)^2 \rho + 2  tf(r)f'(r) T(u,v) +  f(r)^2  p_v \big].
\end{split}    
\end{equation*}
On the other hand,
\begin{equation*}
\begin{split}
     T_{\rm perfect}(\d_t,\d_t) &= \frac{e^{2\Phi(\zeta)}}{t^2 f'(r)^2-f(r)^2} \big[ (\rho_0 + p_0) t^2f'(r)^2 + p_0(-  t^2f'(r) ^2 +  f(r)^2  )\big]\\
     &= \frac{e^{2\Phi(\zeta)}}{t^2 f'(r)^2-f(r)^2} \big[ \rho_0 t^2f'(r)^2 + p_0 f(r)^2  \big].
\end{split}  
\end{equation*}
Therefore
\begin{equation*}
\begin{split}
   & \:\:\:\:\:\:\: T(\d_t,\d_t) -T_{\rm perfect}(\d_t,\d_t)  
   \\
   &=\, \frac{e^{2\Phi(\zeta)}}{t^2 f'(r)^2-f(r)^2} \big[ t^2f'(r)^2 (\rho-\bar\rho) + 2  tf(r)f'(r) T(u,v) +  f(r)^2  (p_v-\bar p) \big].
\end{split}    
\end{equation*}

Combining \eqref{eq-limit}, \eqref{eq-rho origin}, \eqref{eq-limit Tuv}, and \eqref{eq-p origin}, we obtain
\begin{equation*}
     T(\d_t,\d_t) -T_{\rm perfect}(\d_t,\d_t) 
     \,\to\, 0,
\end{equation*}
as we approach the origin $\mc{O}$ within $M_\e$. 
In a similar manner, all other components of $T - T_{\rm perfect}$ in $(t,x,y,z)$-coordinates also converge to 0.
Consequently, assumption (b) in Theorem \ref{thm-main} is satisfied as well. Hence the conclusions of the Theorem \ref{thm-main} hold:
\begin{equation}\label{eq- conclusion}
\rho_0 \,=\, -p_0 \,=\, \frac{3}{4\pi} e^{-2 \Phi(0)}  \Phi'(0) \quad\quad \text{ and } \quad\quad \text{Ric}|_{\mc{O}} \,=\, 8\pi \rho_0\, g|_\mc{O}.
\end{equation}

\noindent\emph{Remark.} We emphasize that we have applied Theorem \ref{thm-main} to the spacetime $M_\e$ and not $M$; this is sufficient for the analysis in the next section.

\subsection{Existence of inflationary eras in the examples}\label{sec: inflation scenario}

In this section we show how the conclusion of our main result, Theorem \ref{thm-main}, proves the existence of inflationary eras for the examples constructed in section \ref{sec: the examples}. 
A majority of the analysis in this section was outlined in 
section 3 of \cite{Ling_remarks_cosmo}. (At the time of writing \cite{Ling_remarks_cosmo}, we had not yet found anisotropic examples of our main theorem which is a main inspiration for writing this paper.)

To gain some familiarity with the problem at hand, let's consider the FLRW setting. Friedmann's second equations is 
\begin{equation}\label{eq: Friedmann}
3\frac{a''(\tau)}{a(\tau)} \,=\, -4\pi \big(\rho(\tau) + 3p(\tau)\big).
\end{equation}
Therefore
\begin{equation}\label{eq: FLRW inflation}
\rho(0) = -p(0) > 0 \quad \Longrightarrow \quad a''(\tau) > 0 \text{ for } \tau \text{ near } \tau = 0.
\end{equation}
The assumption in \eqref{eq: FLRW inflation} is what we mean by ``the cosmological constant appears as an initial condition." It holds for a class of Milne-like spacetimes, see \cite[eq. (1.11)]{Ling_remarks_cosmo}. In fact, our main result, Theorem \ref{thm-main}, is essentially an anisotropoic generalization of this.

In this section, we generalize \eqref{eq: FLRW inflation} to our anisotropic examples. Specifically what we demonstrate is the following. Let $(M,g)$ be the spacetime  defined by equations \eqref{eq: M def 1}, \eqref{eq: M def 2}, and \eqref{eq: M def 3}.  Let $\g(\tau)$ denote a comoving observer in $M$ (i.e., $\g$ is an integral curve of $u$). Here $\tau$ is the proper time of the comoving observer, and we fix it so that $\g(0) = \mc{O}$. We will define a ``generalized scale factor" $\mf{a}(\tau)$ associated with $\g(\tau)$ and show that this generalized scale factor is accelerating, $\mf{a}''(\tau) > 0$, for proper times $\tau$ near $\tau = 0$ (i.e., near the big bang). However, we only prove that some comoving observers experience inflation. Recall that the comoving observers follow the trajectories $t = \mu f(r)$, see \eqref{eq-integral curve}. We find that only those comoving observers with a $\mu$-value above a certain threshold $\mu_{\rm crit}$ experience an inflationary era. This threshold is given by \eqref{eq-mu threshold}; it's completely determined by the functions $f(r)$ and $\Phi(\zeta)$ appearing in the previous section, and hence depends solely on the spacetime metric. 

\medskip
\medskip

\noindent\emph{Remark.} Throughout this section, we have in mind a fixed comoving observer. Since the comoving observers travel along the trajectories $t = \mu f(r)$, we can assume any fixed comoving observer is contained in some $M_\e$ (see \eqref{eq: M epsilon def}) by choosing $\e > 0$ small enough. Therefore the bound \eqref{eq-limit} can be utilized.

\medskip
\medskip


\begin{figure}[h]
\[
 \begin{tikzpicture}[scale = 0.65]

\shadedraw [color = white, bottom color=bello!40!white, top color = white] (10,-2) -- +(45:1.5) arc (45:133:1.5);
\draw [dashed, thick, blue] (10,-2) .. controls (12.8,1) .. (13.3,2.2);
\draw [dashed, thick, blue] (10,-2)  .. controls (7.2,1) .. (6.7,2.2);

\draw [<->,thick] (10,-3.5) -- (10,2.25);
\draw [<->,thick] (5.5,-2) -- (14.5,-2);

\draw (9.65,2.5) node [scale = .75] {$t$};
\draw (14.75, -2.25) node [scale = .75] {$x^i$};


	\draw[color = red] (11.5,2.2) node [scale = .75] {$t=\mu f(r)$};
        \draw [thick, red] (10,-2) .. controls (10.5, -1).. (11,2);
       
        \fill[black] (10,-2) circle(0.1);
        \draw[color = black] (9.75,-2.3) node [scale = .75] {$\mc{O}$};

\draw[fill=none,dashed, thick, black](10,-2) circle (1.5) ;

\end{tikzpicture}
\]
\caption{\small{We prove the existence of inflationary eras along a fixed comoving observer by computing the terms on the right-hand side of the Raychaudhuri equation \eqref{eq-Raychadhuri} in a small neighborhood in $M$ about the origin.}}
\label{fig: inflation}
\end{figure}
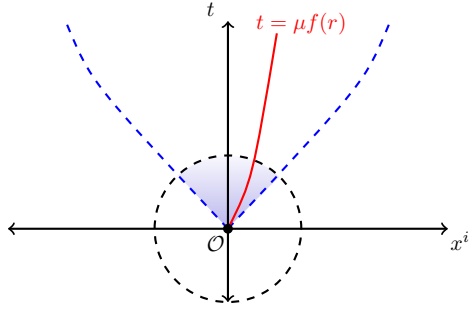

Recall $u$ is given by \eqref{eq: u}. By construction $u$ is orthogonal to the spacelike hypersurfaces of constant $\zeta$. In Figure \ref{fig: comoving observers} (right), one should image that the spacelike hypersurfaces $\tau = \rm{constant}$ are replaced with $\zeta = \rm{constant}$. In the terminology of \cite[p. 359]{ON}, $u$ is ``synchronizable," but it is not necessarily ``proper time synchronizable." The latter occurs if and only if $u$ is geodesic which occurs if and only if $f(r) = r$, see eq. \eqref{eq- acceleration}. (Recall $f(r) = r$ corresponds to a Milne-like spacetime, and we know $u$ is geodesic in this case.)

Set $H = \tfrac13\div u$ so that $H$ coincides with the mean curvature of the spacelike hypersurfaces orthogonal to $u$, i.e., $H$ is one-third the trace of the second fundamental form $K$.\footnote{In the physics literature, $\div u$ is often called the \emph{expansion} $\theta$ of the congruence formed by the integral curves of $u$.} Let $\tau$ denote the proper time of the flow lines of $u$ (i.e., the proper time of the comoving observers). If $c(r)$ denotes the curve $r \mapsto \big(\mu f(r), r, \theta_0, \varphi_0\big)$ along the trajectory $t = \mu f(r)$, then the proper time $\tau$ is simply \begin{equation}\label{eq-proper time}
    \t(r) \,=\, \int_0^r \sqrt{-g\big( c'(s),c'(s)\big)}\,ds \,=\, \int_0^r e^{\Phi(\zeta)} \sqrt{\mu^2 f'(s)^2 -1}\,ds.
\end{equation}
When $c(r)$ is reparameterized by $\tau$, it yields a comoving observer $\g(\tau)$.

Along each comoving observer $\g(\tau)$, we define a \emph{generalized scale factor} $\mathfrak{a}(\tau)$ by\footnote{The generalized scale factor $\mf{a}(\tau)$ is also known as an ``average length scale", see \cite[eq. (2.14)]{Ellis:1971pg}. In the isotropic FLRW setting,  $\mf{a}(\tau)$ is simply the scale factor $a(\tau)$, and the mean curvature $H$ corresponds to the Hubble parameter. It's a happy coincidence that the notation for the mean curvature and the Hubble parameter happen to coincide.} 
\begin{equation}\label{eq- generalized a}
\frac{\mf{a}'}{\mf{a}} \,=\, H. 
\end{equation}

 We have $H(\tau) \approx \frac{1}{\tau}$ for $\tau$ small along each comoving observer $\g(\tau)$, see  eq. \eqref{eq: H limit} below.   Since $\mf{a}(\tau)  = \exp(\int_{\tau_0}^\t H)$ for some arbitrary time $\tau_0$, it follows that 
 \begin{equation}\label{eq: scale factor to 0}
 \mf{a}(\tau) \to 0 \quad \text{ as } \quad \tau \to 0
 \end{equation}
  along each comoving observer. Recall that, for us, the big bang corresponds to the time when the scale factor limits to $0$. Therefore \eqref{eq: scale factor to 0} suggests that the origin point $\mc{O}$ represents the big bang in these models.

For FLRW spacetimes, Friedmann's second equation \eqref{eq: Friedmann} is used to analyze the acceleration of the scale factor. In the anisotropic setting, the generalization of Friedmann's second equation is the Raychaudhuri equation \cite[eq. (4.26)]{Hawking_Ellis},
\begin{equation}\label{eq-Raychadhuri}
    3\frac{\mf{a}''}{\mf{a}} =- \Ric(u,u) - 2\sigma^2 + \div(\nabla_u u).
\end{equation}
(The vorticity term vanishes since $u$ is hypersurface orthogonal.)

Our goal is to compute all the terms on the right-hand side of \eqref{eq-Raychadhuri} for points along a comoving observer near $\mc{O}$ (see Figure \ref{fig: inflation}). First, using the conclusions of Theorem \ref{thm-main} and eq. \eqref{eq- conclusion},  sufficiently close to the origin $\mc{O}$, we have

\begin{equation}\label{eq-ric(u,u)}
    -\Ric(u,u)  \,\approx\, 8\pi \rho_0 \,=\, 6 e^{-2 \Phi(0)}  \Phi'(0).
\end{equation}
The $\approx$ in the above expression is understood in the following way: $-\text{Ric}(u,u)$ can be made arbitrarily close to $8\pi \rho_0$ by choosing points in $M$ arbitrarily close to $\mc{O}$.

The shear term is defined by $2\sigma^2 = \sum_{i,j = 1}^3\sigma(e_i, e_j)\sigma(e_i,e_j)$ where $\{e_1, e_2, e_3\}$ is an orthonormal basis spanning $u^\perp$ and 
\[
\sigma(e_i,e_j) \,=\, K(e_i, e_j) - H \delta_{ij},
\]
where $K(X,Y) = g(\nabla_X u, Y)$ is the second fundamental form of the hypersurfaces orthogonal to $u$. (Recall $H = \frac{1}{3}\text{tr} K$.) Choosing the orthonormal basis $\{v, e_\theta, e_{\varphi}\}$ from the previous section, the only nonvanishing terms for $K$ are
\begin{equation*}
    \begin{split}
        K(v,v) \,&=\, \frac{e^{-\Phi(\zeta)}}{\sqrt{t^2 f'(r)^2-f(r)^2}}\left[  f'(r)  -\frac{t^2 f(r) f'(r) f''(r)}{t^2 f'(r)^2-f(r)^2} +\Phi'(\zeta) \frac{t^2 f'(r)^2-f(r)^2}{f'(r)} \right]\\
        K(e_\theta, e_\theta) \,&=\, K(e_\varphi e_\varphi) \,=\, \frac{e^{-\Phi(\zeta)}}{\sqrt{t^2 f'(r)^2-f(r)^2}}\left[ \frac{f(r)}{r}  +\Phi'(\zeta) \frac{t^2 f'(r)^2-f(r)^2}{f'(r)} \right].
    \end{split}
\end{equation*}
Therefore the mean curvature $H$ is
\begin{equation*}
    3H \,=\, \frac{e^{-\Phi(\zeta)}}{\sqrt{t^2 f'(r)^2-f(r)^2}}\left[  f'(r) + \frac{2f(r)}{r} -\frac{t^2 f(r) f'(r) f''(r)}{t^2 f'(r)^2-f(r)^2}  +3\Phi'(\zeta) \frac{t^2 f'(r)^2-f(r)^2}{f'(r)}   \right].
\end{equation*}

Along $t = \mu f(r)$, we have $H\big|_{t = \mu f(r)} = \frac{e^{-\Phi(0)}}{r\sqrt{\mu^2 -1}} + O(r).$ Using \eqref{eq-proper time}, we reparameterize in terms of $\tau$ giving
\begin{equation}\label{eq: H limit}
H|_{\g(\tau)} \,=\, \frac{1}{\tau} + o(1).
\end{equation}

Direct computation shows that 
\begin{equation}\label{eq- shear}
    2\sigma^2 \,=\, \frac{2e^{-2\Phi(\zeta)}}{3[t^2 f'(r)^2-f(r)^2]}\left[  f'(r) - \frac{f(r)}{r} -\frac{t^2 f(r) f'(r) f''(r)}{t^2 f'(r)^2-f(r)^2} \right] ^2.
\end{equation}
For small $r$, we have $f'(r) - \frac{f(r)}{r} = O(r^2)$ which combined with \eqref{eq-limit} yields
\begin{equation}\label{eq- shear to 0}
    2\sigma^2 \,\to\, 0.
\end{equation}
 In other words $2\sigma^2$ extends continuously to $M_\e \cup \{\mc{O}\}$  and takes on the value 0 at $\mc{O}$. Geometrically, this ``isotropization" effect is a consequence of the $u$-orthogonal hypersurfaces becoming more hyperbolic as we approach the origin $\mc{O}$.


The last term in \eqref{eq-Raychadhuri} to compute is $\div(\nabla_u u)$. We have
\begin{equation}\label{eq- acceleration}
    \nabla_u u = 
    -\frac{t f(r)^2 f''(r) e^{-2\Phi(\zeta)} }{( t^2f'(r)^2 - f(r)^2 )^2} \Big( f(r) \partial_t +t  f'(r) \partial_r\Big).
\end{equation}
Hence
\begin{equation*}
\begin{split}
    \div(\nabla_u u) = -&\frac{ f(r) e^{-2\Phi(\zeta)} }{( t^2f'(r)^2 - f(r)^2 )^2} \bigg[  2t^2 f'(r)^2 f''(r) + t^2 f(r) f''(r)^2 + t^2 f(r) f'(r) f'''(r)   \\ & \qquad\quad\quad\left.+ f(r)^2 f''(r) + \frac{2t^2 f(r) f'(r) f''(r)}{r}  - \frac{4 t^4 f(r) f'(r)^2 f''(r)^2}{ t^2f'(r)^2 - f(r)^2 }   \right].
\end{split}
\end{equation*}
Evaluating along $t = \mu f(r)$ and taking the limit $r \to 0$, we find
\begin{equation}\label{eq-div along co obs}
\begin{split}
    \div(\nabla_u u)\big|_{t = \mu f(r)} \,=\, & - e^{-2\Phi(0)}\left[\frac{  f'''(0)( 5\mu^2  + 1) }{( \mu^2 - 1 )^2}+ {O(r)}\right] .
\end{split}
\end{equation}

Using \eqref{eq-ric(u,u)}, \eqref{eq- shear to 0}, and \eqref{eq-div along co obs}, the Raychaudhuri equation \eqref{eq-Raychadhuri}, for points along the comoving observer sufficiently close to the origin $\mc{O}$, becomes
\begin{equation}\label{eq-raych at origin}
    3\frac{\mf{a}''}{\mf{a}}\bigg|_{t = \mu f(r)} \,\approx\, 6e^{-2 \Phi(0)}  \left[ \Phi'(0)  -\frac{\tfrac{1}{6}f^{'''}(0)  (  5\mu^2  + 1) }{( \mu^2 - 1 )^2} \right].
\end{equation}
Similar to \eqref{eq-ric(u,u)}, the $\approx$ symbol in the above expression is understood in the following way: $3(\mf{a}''/\mf{a})$ can be made arbitrarily close to the right-hand side of \eqref{eq-raych at origin} by choosing points along $t = \mu f(r)$ that are sufficiently close to the origin $\mc{O}$.


From \eqref{eq-raych at origin} we can determine which comoving observers experience an inflationary era, i.e., which comoving observers experience $\mf{a}''(\tau) > 0$ arbitrarily close to  $\tau = 0$.  Assuming $\Phi'(0) > 0$ (which is equivalent to $\wt{\rho}(\mc{O}) > 0$), it's precisely those comoving observers with $\mu$-values satisfying
\begin{equation}\label{eq-mu threshold}
    \mu\,>\, \mu_{\rm crit} \,:=\, \sqrt{\frac{12\Phi'(0) + 5f'''(0) +\sqrt{144\Phi'(0)f'''(0) +25  f'''(0)^2}}{12\Phi'(0)}} \,.
\end{equation}
Moreover, we see that if $f'''(0) = 0$ and $\Phi'(0) > 0$, then all the comoving observers experience an inflationary era. This reproduces the results for Milne-like spacetimes, see \eqref{eq: FLRW inflation}.

\subsection{Remarks on proving anisotropy}

In this section we show that the examples constructed in section \ref{sec: the examples} are generally anisotropic. Although this is heuristically evident, a formal mathematical proof is not immediately clear.

First, the definition of an ``isotropic spacetime" is not consistent throughout the literature. See  \cite{Rodrigo_cosmo} and \cite{Sanchez_cosmo} and references therein. We will adopt the definition in \cite[Ch. 12]{ON} since, as discussed in \cite{Rodrigo_cosmo}, this definition is the optimal one as it implies that the spacetime is isometric to a subset of an FLRW spacetime, see \cite[Prop. 12.6]{ON} and \cite[Thm. 2.1]{Rodrigo_cosmo}.

Therefore any spacetime that is not isometric to a subset of an FLRW model is anisotropic according to \cite{ON}. For the examples constructed in section \ref{sec: the examples}, if $f(r) = r$ then they are isometric to a subclass of Milne-like spacetimes which are a subclass of $k = -1$ FLRW models, and hence they are isotropic. 
 Moreover, regardless of the form of $f(r)$, if the conformal factor is identically $1$, then the spacetime is isometric to a subset of Minkowski spacetime which is clearly isotropic. (This shows that it is not sufficient to simply recognize that the shear term \eqref{eq- shear} is nonzero. However, in this case, the vector field defining the comoving observers changes.) This suggests that if $f(r)$ is not identically $r$ and the conformal factor is not constant, then the resulting spacetime is not a subset of an FLRW spacetime and hence is anisotropic. We believe such a statement can be proven rigorously. However, in this section, we will content ourselves with the following algorithm: Pick functions $f(r)$ and $\Phi(\zeta)$. The steps below show how to verify that the corresponding spacetime $(M,g)$ from section \ref{sec: the examples} is anisotropic.
 
 Seeking a contradiction, suppose $(M,g)$ is in fact isometric to a subset of an FLRW spacetime. Since FLRW spacetimes satisfy the Einstein equations with a perfect fluid \cite[Thm. 12.11]{ON}, there is a unit future directed timelike vector field $\wt{u}$ on $M$ such that $T$ is a perfect fluid with respect to $\wt{u}$. There exist functions $a$, $b$, $c$, $d$ such that 
\[
\wt{u} \,=\, au + bv + ce_\theta + de_\varphi,
\]
 where $\{u, v, e_\theta, e_{\varphi}\}$ is the orthonormal frame constructed in section \ref{sec: the examples}. Consider the unit spacelike vectors orthogonal to $\wt{u}$
\[
\wt{v} \,=\, \frac{bu + av}{a^2 - b^2}, \quad\quad \wt{e}_\theta \,=\, \frac{cu + ae_\theta}{a^2 - c^2}, \quad\quad \wt{e}_\varphi \,=\, \frac{du + ae_{\varphi}}{a^2 - d^2}.
\]
From section \ref{sec: the examples}, we know how $T$ acts on the orthonormal frame $\{u, v, e_\theta, e_\varphi\}$, and so we know how $T$ acts on $\{\wt{u}, \wt{v}, \wt{e}_\theta, \wt{e}_\varphi\}$.

Fix a point $p_0 \in M$ given by $(t_0, r_0, \theta_0, \varphi_0)$.  At $p_0$, the following equations set up an overdetermined system for $(a,b,c,d)$ at $p_0$.
\begin{align*}
-a^2 + b^2 + c^2 + d^2 \,&=\, -1
\\
T(\wt{u}, \wt{v}) \,&=\, 0
\\
T(\wt{u}, \wt{e}_\theta) \,&=\, 0
\\
T(\wt{u}, \wt{e}_\varphi) \,&=\,0
\\
T(\wt{v}, \wt{v}) \,=\, T(\wt{e}_\theta, \wt{e}_\theta) \,&=\, T(\wt{e}_\varphi, \wt{e}_\varphi).
\end{align*}
For most choices of $f(r)$ and $\Phi(\zeta)$, this system does not have any solutions, giving a contradiction. However, even if there are solutions, one can still obtain a contradiction by other means, e.g., showing that the orthogonal subspace to $\wt{u}$ does not have constant sectional curvature. Lastly, we remark that the point $p_0$  must lie away from $r = 0$. Indeed, points along $r = 0$ will past the above tests. This is due to the spacetime being spherically symmetric and hence spatially isotropic precisely at points along $r = 0$.

%

We remark that if $\Phi$ is constant or $f(r) = r$, then $(M,g)$ passes the above tests. In the first case, the metric is homothetic to the Minkowski metric (and hence isometric to a subset of a $k = 0$ FLRW spacetime), and in the second case, the spacetime is given by a Milne-like spacetime and hence is isometric to a $k = - 1$ FLRW spacetime.

\section{Summary and outlook}

The inflationary scenario has become the current paradigm of early universe cosmology. Roughly, it states that scale factor underwent a brief but dramatic period of acceleration after the big bang but before the radiation dominated era. Although inflationary theory has many successes (e.g., solutions to the horizon and flatness problems along with providing a framework for generating the seeds of large-scale structures in our universe), it does not carry the status of an established physical theory. In this work, we provide mathematical support for the inflationary scenario by showing that a class of anisotropic spacetimes experience inflationary eras after the big bang. 

Our main result, Theorem \ref{thm-main}, says that if the universe began with special initial conditions at the big bang, then the energy-momentum tensor was dominated by a cosmological constant at the big bang. These special initial conditions are (1) the existence of an origin point $\mc{O}$ for a unit timelike vector field $u$ (whose integral curves represent the comoving observers in the universe) and (2) the energy-momentum tensor approaches a perfect fluid at $\mc{O}$. An informal discussion of these special initial conditions is given in section \ref{sec: geom initial}.

In section \ref{sec: the examples}, we construct anisotropic spacetimes which satisfy the hypotheses of Theorem \ref{thm-main}. These examples can be thought of as ``quasi Milne-like spacetimes." In section \ref{sec: inflation scenario}, we define a generalized scale factor $\mf{a}(\tau)$ along each comoving observer ($\tau$ denotes the proper time of the comoving observer), and we show that $\mf{a}(\tau) \to 0$ as $\tau \to 0$, see \eqref{eq: scale factor to 0}. Consequently, we associate $\tau = 0$ (and hence also the origin point $\mc{O}$) with the big bang. Lastly, we describe which comoving observers experience inflation, $\mf{a}''(\tau) > 0$, immediately after the big bang $\tau = 0$. See equations \eqref{eq-raych at origin} and \eqref{eq-mu threshold}.

Our examples exhibit isotropization towards the past and, in fact, are perfectly isotropic at the big bang $\mc{O}$, see \eqref{eq- shear to 0}. Our isotropization-towards-the-past result is consistent with a universe starting from special initial conditions. This is unlike results related to the cosmic no-hair conjecture (see  Wald's original paper \cite{Wald_cosmic_no_hair} or some more recent work, e.g., \cite{Azhar_Kaiser}), where isotropization occurs towards the future.

A limitation of our approach is that we only show accelerated expansion immediately after the big bang. For example reheating does not appear in our analysis. For an analysis of the physics after the accelerated expansion, our geometrical initial conditions should be supplemented with, for example, appropriate scalar field matter models.

We believe that differential geometry (and geometric analysis in particular) has a role to play in the investigation of initial conditions for the big bang. The work presented in this paper should be thought of as a ``proof of concept" of this proposal. Our work can be generalized in many ways. In particular, although our examples are not necessarily isotropic, they are still spherically symmetric. So a natural generalization is to reproduce the analysis in sections \ref{sec: the examples} and \ref{sec: inflation scenario} with non-spherically symmetric spacetimes. Also, our examples are anisotropic versions of $k = -1$ FLRW spacetimes. What about $k = 0$ FLRW spacetimes? In this case one would want to apply \cite[Thm. 5.2]{GLQ} or a suitable generalization thereof. Lastly, 
it remains to be seen if the results in \cite{gao2021inverse} can be used to generate comoving observers with an origin point $\mc{O}$.

\section*{Acknowledgments} 
Eric Ling was supported by Carlsberg Foundation CF21-0680 and Danmarks Grundforskningsfond CPH-GEOTOP-DNRF151. Annachiara Piubello was supported by the DFG Project ME 3816/3-1, part of the SPP2026. We thank Jerome Quintin for helpful comments on an earlier draft and are grateful to the Minkowski institute where this project began to take shape.

%
%
%
%

\let\oldbibliography\thebibliography
\renewcommand{\thebibliography}[1]{
  \oldbibliography{#1}
  \setlength{\itemsep}{1pt}
}


\newpage

\begin{thebibliography}{10}

\bibitem{albrecht2003cosmic}
Andreas Albrecht, \emph{Cosmic inflation and the arrow of time}, 2003,
  arXiv:astro-ph/0210527v3.

\bibitem{Rodrigo_cosmo}
Rodrigo Avalos, \emph{{On the rigidity of cosmological space-times}}, Letters
  of Math Phys. \textbf{113} (2023), no.~98.

\bibitem{Azhar_Kaiser}
Ferez Azhar and David I. Kaiser, \emph{Flows into de Sitter space from anisotropic initial conditions: An effective field theory approach}, Phys. Rev. D \textbf{107} (2023), no. 4, 043506.

\bibitem{Brandenberger:2016uzh}
Robert Brandenberger, \emph{{Initial conditions for inflation \textemdash{} A
  short review}}, Int. J. Mod. Phys. D \textbf{26} (2016), no.~01, 1740002.
  
\bibitem{Brandenberger_bouncing}
Robert Brandenberger and Patrick Peter, \emph{Bouncing Cosmologies: Progress and Problems}, Found. Phys. \textbf{47} (2017), no. 6, 797-850.

\bibitem{carroll2004spontaneous}
Sean~M. Carroll and Jennifer Chen, \emph{Spontaneous inflation and the origin
  of the arrow of time}, 2004, arXiv:hep-th/0410270.

\bibitem{PhysRevD.21.3305}
Sidney Coleman and Frank De~Luccia, \emph{Gravitational effects on and of
  vacuum decay}, Phys. Rev. D \textbf{21} (1980), 3305--3315.

\bibitem{Planck:2018jri}
Planck Collaboration, \emph{{Planck 2018 results. X. Constraints on
  inflation}}, Astron. Astrophys. \textbf{641} (2020), A10.

\bibitem{Ellis:1971pg}
George F.R. Ellis, \emph{{Relativistic cosmology}}, Proc. Int. Sch. Phys. Fermi
  \textbf{47} (1971), 104--182.

\bibitem{Galloway_Ling_inextend}
Gregory Galloway and Eric Ling, \emph{{Some Remarks on the
  $C^0$-(in)extendibility of Spacetimes}}, Annales Henri Poincare \textbf{18}
  (2017), no.~10, 3427--3447.

\bibitem{gao2021inverse}
Ya~Gao and Jing Mao, \emph{Inverse mean curvature flow for spacelike graphic
  hypersurfaces with boundary in {L}orentz-{M}inkowski space
  $\mathbb{R}^{n+1}_{1}$}, 2021, arXiv:2104.10600v5.

\bibitem{GLQ}
Ghazal Geshnizjani, Eric Ling, and Jerome Quintin, \emph{On the initial
  singularity and extendibility of flat quasi-de {S}itter spacetimes}, Journal
  of High Energy Physics \textbf{2023} (2023), no.~10.

\bibitem{Guth:1980zm}
Alan~H. Guth, \emph{{The Inflationary Universe: A Possible Solution to the
  Horizon and Flatness Problems}}, Phys. Rev. D \textbf{23} (1981), 347--356.

\bibitem{Guth_2014}
Alan~H. Guth, David~I. Kaiser, and Yasunori Nomura, \emph{Inflationary paradigm
  after planck 2013}, Physics Letters B \textbf{733} (2014), 112–119.

\bibitem{Hawking_Ellis}
Stephen~W. Hawking and George F.~R. Ellis, \emph{{The Large Scale Structure of
  Space-Time}}, Cambridge Monographs on Mathematical Physics, Cambridge
  University Press, 2023.

\bibitem{Ijjas:2013vea}
Anna Ijjas, Paul~J. Steinhardt, and Abraham Loeb, \emph{{Inflationary paradigm
  in trouble after Planck2013}}, Phys. Lett. B \textbf{723} (2013), 261--266.

\bibitem{Linde_2018}
Andrei Linde, \emph{On the problem of initial conditions for inflation},
  Foundations of Physics \textbf{48} (2018), no.~10, 1246–1260.

\bibitem{Linde:1981mu}
Andrei~D. Linde, \emph{{A New Inflationary Universe Scenario: A Possible
  Solution of the Horizon, Flatness, Homogeneity, Isotropy and Primordial
  Monopole Problems}}, Phys. Lett. B \textbf{108} (1982), 389--393.

\bibitem{Ling_causal_theory}
Eric Ling, \emph{{Aspects of $C^0$ causal theory}}, Gen. Rel. Grav. \textbf{52}
  (2020), no.~6, 57.

\bibitem{Ling_coord_singu}
Eric Ling, \emph{{The Big Bang is a Coordinate Singularity for $k = -1$
  Inflationary FLRW Spacetimes}}, Found. Phys. \textbf{50} (2020), no.~5,
  385--428.

\bibitem{Ling_remarks_cosmo}
Eric Ling, \emph{{Remarks on the cosmological constant appearing as an initial
  condition for Milne-like spacetimes}}, Gen. Rel. Grav. \textbf{54} (2022),
  no.~7, 68.

\bibitem{Ling_Piubello}
Eric Ling and Annachiara Piubello, \emph{{On the asymptotic assumptions for
  Milne-like spacetimes}}, Gen. Rel. Grav. \textbf{55} (2023), no.~4, 53.

\bibitem{Martin:2013tda}
Jerome Martin, Christophe Ringeval, and Vincent Vennin, \emph{{Encyclop\ae{}dia
  Inflationaris}}, Phys. Dark Univ. \textbf{5-6} (2014), 75--235.

\bibitem{Mukhanov:1981xt}
Viatcheslav~F. Mukhanov and G.~V. Chibisov, \emph{{Quantum Fluctuations and a
  Nonsingular Universe}}, JETP Lett. \textbf{33} (1981), 532--535.

\bibitem{Nomura:2021lzz}
Kimihiro Nomura and Daisuke Yoshida, \emph{{Past extendibility and initial
  singularity in Friedmann-Lema\^\i{}tre-Robertson-Walker and Bianchi I
  spacetimes}}, JCAP \textbf{07} (2021), 047.

\bibitem{ON}
Barrett O'Neill, \emph{Semi-{R}iemannian geometry}, Pure and Applied Mathematics,
  vol. 103, Academic Press Inc. [Harcourt Brace Jovanovich Publishers], New
  York, 1983.

\bibitem{Penrose_time}
Roger Penrose, \emph{{Singularities and Time Asymmetry}}, pp.~581--638, 1980,
  {General Relativity}: {An Einstein Centenary Survey}.

\bibitem{Sanchez_cosmo}
Miguel S\'{a}nchez, \emph{{A class of cosmological models with spatially
  constant sign-changing curvature}}, Portugaliae Mathematica \textbf{80}
  (2023), no.~3/4.

\bibitem{Starobinsky:1980te}
Alexei~A. Starobinsky, \emph{{A New Type of Isotropic Cosmological Models
  Without Singularity}}, Phys. Lett. B \textbf{91} (1980), 99--102.
  
\bibitem{Wald_cosmic_no_hair}
Robert M. Wald, \emph{Asymptotic behavior of homogeneous cosmological models in the presence of a positive cosmological constant}, Phys. Rev. D \textbf{28} (1983), 2118--2120.

\end{thebibliography}

\end{document}